\documentclass[12pt,preprint]{aastex}

\slugcomment{Submitted to the Astrophysical Journal}
\shorttitle{Accelerated Electrons in Cas A } \shortauthors{Laming}
\begin{document}
\title{Accelerated Electrons in Cassiopeia A: Thermal and Electromagnetic
Effects}

\author{J. Martin Laming}
\affil{E. O. Hulburt Center for Space Research, Naval Research
Laboratory,
       Code 7674L, Washington DC 20375}
\email{jlaming@ssd5.nrl.navy.mil}

\begin{abstract}
We consider in more detail a model previously proposed for the hard X-ray
($>10$ keV) emission observed from the supernova remnant Cas A, whereby
electrons are accelerated by lower-hybrid waves and radiate bremsstrahlung.
We consider both cold and thermal plasma limits of the modified two-stream
instability that generates the lower-hybrid waves, and by studying time dependent
ionization balance for various components of the Cas A ejecta and shocked
circumstellar medium, find locations within the shell where one or other of the
instabilities may occur. Either instability can be effective,
with the cold plasma instability imposing fewer constraints on the shocked reflected
ion population responsible for exciting the waves.
The instability must be located in the ejecta shocked at the earliest
times and therefore closest to the contact discontinuity where magnetic fields are
expected to be the strongest. The energy deposited in this ejecta by
collisions between accelerated and ambient electrons is broadly consistent with
that required to reheat this ejecta to the observed temperature of $\sim 4\times 10^7$ K.

\end{abstract}



\section{Introduction}
The supernova remnant Cassiopeia A has long intrigued astonomers across
the entire electromagnetic spectrum. Long known as the brightest radio source in the
northern sky \citep{reber44,ryle48}, it has also recently been detected in
TeV $\gamma$ rays \citep{aharonian01},
and is most likely the remnant of a putative supernova
observed as a 6th magnitude star by Flamsteed in 1680. The discovery of
hard X-ray continuum emission extending out to 100 keV has stimulated
interest in electron acceleration in its shock waves. As discussed in a
recent paper \citep{laming01}, the original interpretation of this
emission as synchrotron radiation from highly relativistic electrons
\citep{allen97}
deserves further study. Unlike in the cases of SN 1006 \citep{koyama95},
G347.3-0.5 \citep{slane99,koyama97}, and now also G266.1-1.2 \citep{slane00},
where the X-ray continuum is strong enough to completely mask the line
emission that should also be present, the continuum in Cas A is merely
a high energy ``tail'' to the otherwise Maxwellian thermal bremsstrahlung
continuum, as shown in Figure 1. (Figure 1 also shows some models discussed in detail
below). It is thus possible that such emission could be the result of
electrons accelerated by collisionless processes other than first order
Fermi acceleration at shock fronts.

Such a scenario is possible because the shock transition occurs on a length
scale much shorter than particle mean free paths to Coulomb collisions.
Assuming that the gas pressure $>>$ magnetic pressure in the
downstream plasma, the temperature $T_a$ reached by particle
species $a$ with mass $m_a$ is given by conservation of energy,
momentum and particles across the shock discontinuity by
\begin{equation}\label{eqn1.1}
k_{\rm B}T_a={3\over 16}m_av_s^2
\end{equation}
for shock velocity $v_s$. Thus in the limit of true collisionless
plasma, the particles will have temperatures proportional to their
masses, i.e. protons will have a temperature $m_i/m_e = 1836$
times that of the electrons. Obviously in realistic conditions
Coulomb equilibration will take place, with rate
\citep[][see also Appendix B]{spitzer78}
\begin{equation}\label{eqn2.2}
{d\Delta T\over dt}=-0.13Z^2n_e{\Delta T\over AT_e^{3/2}}
\end{equation}
in c.g.s. units, where $\Delta T=T_i-T_e$, and the ions have
charge $Z$ and atomic mass $A$. However collisionless processes
involving the generation of plasma waves at the shock front and
the associated electron heating by Landau damping or stochastic processes
can operate on
a much faster timescale. Recent simulations of such processes are given
by \citet{dieckmann00} and \citet{shimada00} for quasi perpendicular
shocks (magnetic field in the plane of the shock front) and by
\citet{bykov99} for a quasi parallel shock (magnetic field parallel
to shock velocity vector), where shock heated electron distributions
characterized by Maxwellians at low energies going over into power
law distributions at high energies are frequently seen.

Although these results appear qualitatively consistent with the observed
X-ray spectra of Cas A, the situation is complicated by the fact that optical
\citep{ghavamian99} and UV \citep{laming96} diagnostics for collisionless
electron heating at fast shocks (admittedly in other supernova remnants,
Tycho and SN 1006) give essentially no electron heating
beyond that implied by the shock jump conditions, equation (\ref{eqn1.1}). Similar
results have been found in X-ray spectra for E0102.2-7219 \citep{hughes00},
SN 1987A \citep{burrows00}, and SN 1993J \citep{fransson96}.
\citet{laming98,laming00,laming01} speculated that the extrapolation involved from
the modest Mach numbers (of order 1-10) of the simulated shocks to the
very high ($>100$) Mach numbers for forward and reverse shocks in supernova
remnants might not be valid. The reasoning behind this is that the higher
Mach number shocks become progressively more turbulent
\citep[c.f.][]{tidman71,tokar86,cargill88}. Although this appears
to have no effect on the simulations, these have only one spatial dimension. In
two or three dimensions the conjecture is that the shock reflected ions, while
still quasi-monoenergetic, no longer behave as a ``beam'', now having a much larger
distribution in pitch angle. The consequence of this is that the preshock instabilities
heating the electrons are now no longer reactive in nature but kinetic, and have
growth rates low enough that insignificant wave growth and hence electron heating
can occur before the preshock plasma is overrun by the shock front itself. Reactive
and kinetic instabilities are discussed more fully in Appendix A.1.

For this reason \citet{laming01} investigated a slightly different mechanism;
the excitation of lower hybrid waves and their consequent acceleration
of electrons as a result of secondary shocks passing through the presumed
region of strong magnetic field in Cas A.
Lower hybrid waves are electrostatic ion oscillations, which can occur in
magnetic fields strong enough that
the electron gyroradius is smaller
than the lower hybrid wavelength. Due to this necessary magnetization of the electrons,
the waves propagate preferentially across magnetic field lines. The
parallel component of the wavevector $k_{||}/k < \omega _{pi}/\omega _{pe}$. Since
$\omega /k_{\perp} \sim \left(m_e/m_i\right) ^{1/2}\omega /k_{||}$ the wave can simultaneously be in
resonance with ions moving across the magnetic field and electrons moving along magnetic
field lines, which facilitates collisionless energy exchange between ions and electrons
on timescales much faster than that associated with Coulomb collisions. A cold plasma
theory for lower-hybrid waves is given in the Appendix of \citet{laming01}.
They have also been previously
discussed in connection with cometary X-ray emission, supernova remnant shock waves
and Advection Dominated Accretion Flows (ADAFS)
\citep{vaisberg83,krasno85,bingham97,shapiro99,bingham00,begelman88}, and observed
in situ together with accelerated electrons at Halley's comet \citep{gringauz86,klimov86}.

Lower hybrid waves are especially attractive to consider
due to the strong magnetic fields inferred to exist in Cas A \citep[e.g.][]{longair94},
near the contact discontinuity
between ambient circumstellar material shocked by the blast wave and supernova
ejecta heated by the reverse shock. The situation is shown schematically in Figure 2.
The explosion drives a blast wave into the surrounding circumstellar medium. As this
begins to sweep up more plasma behind it and decelerates, a reverse shock develops which
moves back into the ejecta (in a Lagrangian sense; it is still expanding in Eulerian
coordinates), with a contact discontinuity between the reverse shocked ejecta and
forward shocked circumstellar medium. The dense shocked ejecta being decelerated by the
much less dense shocked circumstellar medium is subject to a Rayleigh-Taylor instability,
which is the likely mechanism for the magnetic field amplification. Also shown in Figure
2 are two of the various classes of knots or clumps of material in Cas A
that are particularly conspicuous in optical observations. Quasi-stationary floculli
(QSFs) are seen ahead of the forward shock and fast moving knots (FMKs) ahead of the
reverse shock. As these shocks run into the density contrasts represented by these
structures, they split into a transmitted shock that propagates through the clump
and eventually destroys it, and a reflected shock that moves back towards the contact
discontinuity. It is these reflected shocks that in our model generate the lower hybrid
waves and accelerated electrons in the high magnetic field region.

In this paper we extend the analysis of \citet{laming01} to more realistic
thermal plasma conditions in Cas A. These are taken from time dependent
ionization balance calculations based on analytic expressions for the hydrodynamics
from \citet{mckee95}, and updated in \citet{truelove99}.
Section 2 develops the plasma theory for lower hybrid
waves in thermal conditions. Section 3 describes the ionization balance
calculations with and without the extra collisionless electron heating. Since
the mechanism depends on having strong magnetic fields in the shell of Cas A,
amplified presumably from seed fields by a Rayleigh-Taylor instability, the history
of which over the lifetime of the remnant is presently unknown, it is not
possible to calculate definitively the hard X-ray spectrum. However if this
model is basically correct, a number of inferences can be made about the nature
of the plasma in Cas A, and these are discussed in section 4, followed by
conclusions in section 5. Many of the mathematical details in section 2 and 3 are given
in the Appendices.

\section{Cas A and Lower Hybrid Waves Revisited}
In a previous paper \citep{laming01} a theory of the growth of lower-hybrid
waves was described in the cold plasma approximation (i.e. where the wave
frequency $\omega >>\sqrt{2}
kv_i$ and the electron cyclotron frequency
$\Omega _e >>k_{\perp}v_e$; $v_e$ and $v_i$ are the electron and ion thermal
velocities respectively), where ions reflected from
shocks propagating throughout the shell of Cas A generate waves through a
modified two stream instability at a perpendicular shock. In this case, waves
moving at angle $\theta$ to the magnetic field direction
given by $\cos\theta =\omega _{pi}/\omega _{pe}$ are preferentially excited,
where $\omega _{pi}$ and $\omega _{pe}$ are the ion and electron plasma frequencies
respectively. This
is because waves at this angle have a group velocity away from the shock equal
to the shock velocity itself \citep{mcclements97}, and hence stay in contact
with the shock reflected ions for arbitrary lengths of time.
This allows large wave amplitudes to develop,
even though the growth rates may be small. These waves are excited by ions
{\em returning} to the shock front with velocity $2v_s$ directed along to shock
velocity vector. As the shock becomes quasi-parallel this
feature disappears, since the wave group velocity is no longer in the same direction
as the shock velocity. The generalization of this theory for
thermal electrons and ions is given in Appendix A. Where $\omega >>\sqrt{2}kv_i$,
it is found that outgoing ions can now excite the necessary waves, propagating
at direction cosines essentially the same as those found in the cold plasma limit
for the returning ions,
i.e. $x^2\sim \omega _{pi}^2/\omega _{pe}^2$. In the opposite case, $\omega <<
\sqrt{2}kv_i$, again only returning ions can excite the waves, but for a given
wavevector do so at a variety of small direction cosines.

In the hot ion limit ($\omega << \sqrt{2}kv_i$)
waves can be generated with a range of direction cosines
and frequencies, so we now proceed to determine the wave direction
cosine and frequency at which the net growth rate is maximized in this limit.
The growth rate due to a Maxwellian ion distribution
moving with bulk velocity $\vec{U}$ in the limit $\omega << \sqrt{2}kv_i$
is given by a generalization of equation (A12) of \citet{laming01}, which we
reproduce here,
\begin{equation}
\gamma ={1\over
2}\sqrt{\pi\over 2}\omega _{pi}^{\prime ~2} \left(\omega\over
kv_i\right)^3 \left({\omega _{pe}^2Ik_{||}^2/k^2\over 1+\omega _{pe}^2/k^2c^2}
\right)^{-1}\left(\vec{k}\cdot\vec{U}-\omega\right)
\exp\left[ -\left(\omega -\vec{k}\cdot\vec{U}\right)^2/2k^2v_i^2\right],
\end{equation}
where $I={m_e\over k_{\rm B}T_e}\int _0^{+\infty}  J_0^2\left(k_{\perp}v_{\perp}
\over\Omega _e\right)\exp\left(-m_ev_{\perp}^2
\over 2T_e\right)v_{\perp}dv_{\perp}$, $v_e^2=k_{\rm B}T_e/m_e$, $v_i^2=k_{\rm B}T_i/m_i$,
$J_0\left(k_{\perp}v_{\perp}
\over\Omega _e\right)$ is the Bessel function of the first kind of order zero and
$\Omega _e$ is the electron gyrofrequency. The wave frequency
in this limit is given by equation (A7) with $\phi = \omega ^2/k^2v_i^2+...$
\begin{equation}
\omega ^2={\omega _{pe}^2\left(k_{||}/k\right)^2I/\left(1+\omega _{pe}^2/k^2c^2\right)\over
1+\omega _{pi}^2/k^2v_i^2 +\left(1-I\right)\omega _{pe}^2/k^2v_e^2}.
\end{equation}
The growth rate must exceed the Landau damping rate, $\gamma _{LD}$,
of the wave due to thermal
ions, given by equation (3) with $\vec{k}\cdot\vec{U}=0$.

Since $\omega <<\sqrt{2}kv_i$, we may put $\vec{k}\cdot\vec{U}>>\omega$ and find $\omega$
for which $\gamma -\gamma _{LD}$ is maximized.
Substituting for $k_{||}/k$ in terms of $\omega $ in the expression for
$\gamma -\gamma _{LD}$, we find
\begin{equation}
{\partial\left(\gamma -\gamma _{LD}\right)\over\partial\omega }
=\gamma\left({\vec{k}\cdot\vec{U}-2\omega\over
\omega\vec{k}\cdot\vec{U}-\omega ^2}-{\omega -\vec{k}\cdot\vec{U}\over k^2v_i^2}\right)
-\gamma _{LD}\left({2\over\omega}-{\omega\over k^2v_i^2}\right).
\end{equation}
Setting this equal to zero we find maxima or minima of $\omega $ given by the cubic
equation where $\vec{k}\cdot\vec{U}=-kv_s/\alpha$, with $v_s$ the shock velocity and
$\alpha = 1/\left(2\cos\beta\right)$, where $\beta $ is the angle between $\partial\omega
/\partial\vec{k_{\perp}}$ and $U$;
\begin{eqnarray}
\nonumber &\omega ^3\left[1+{n_i^{\prime}\over n_i}\exp\left(-M^2/\alpha ^2\right)\right]
+\omega ^2\left[{2kv_s\over\alpha }
{n_i^{\prime}\over n_i}\exp\left(-M^2/\alpha ^2\right)\right]\\+
&\omega\left[\left({k^2v_s^2\over\alpha ^2}-2k^2v_i^2\right){n_i^{\prime}\over n_i}\exp
\left(-M^2/\alpha ^2\right)-2k^2v_i^2)\right] -
{kv_s\over\alpha }k^2v_i^2{n_i^{\prime}\over n_i}\exp\left(-M^2/\alpha ^2\right)=0
\end{eqnarray}
where $M=v_s/\left(\sqrt{2}v_i\right)$, the Mach number divided by $\sqrt{2}$.
For $-kv_s/\alpha >> kv_i$ this has two solutions at $\omega = \sqrt{2}kv_i$
corresponding to minima and a maximum
at $\omega =0$. Looking for more exact solutions near $\omega =0$ we
drop the term in $\omega ^3$ and solve the resulting quadratic equation to get
\begin{equation}
\omega =-\alpha{n_i^{\prime}\over n_i}\sqrt{2}kv_i{\left(M^2/\alpha ^2+1\right)^2\over
8M}\exp\left(-M^2/\alpha ^2\right).
\end{equation}
The value of $x$ is then given by
\begin{equation}
x=-\alpha{n_i^{\prime}\over n_i}{\sqrt{2}kv_i\left(M^2/\alpha ^2+1\right)^2
\exp\left(-M^2/\alpha ^2\right)\over 8M
\omega _{pe\left(EM\right)}\sqrt{I}}\left[1+{\omega _{pi}^2\over k^2v_i^2}
+{\omega _{pe}^2\over k^2v_e^2}\left(1-I\right)\right]^{1/2}
\end{equation} and the growth rate by
\begin{equation}
\gamma -\gamma _{LD}=\sqrt{\pi\over 2}{n_i^{\prime}
\over n_i}{\left(M^2/\alpha ^2+1\right)^2\over 8M}{\omega _{pi}^{\prime ~2}kv_i
\exp\left(-2M^2/\alpha ^2\right)\over k^2v_i^2 +\omega _{pi}^2
+\left(1-I\right)\omega _{pe}^2v_i^2/v_e^2}\left[M-
{\alpha ^2\left(M^2/\alpha ^2+1\right)^2\over
8M}\right].
\end{equation}
It is shown in the Appendix that under a broad range of conditions, $-1<\alpha <-1/2$.
Substituting equation (7) into equation(3) is it easy to see that the growth rate
is maximized as $\alpha\rightarrow -1$, and that waves grow fastest for the
maximum allowable $k$.
In realistic conditions $k_{max}\sim \Omega _e/v_e$, since the electron gyroradius
must be shorter than the wavelength for lower hybrid waves. Then for $M=1$ the growth rate
is $\sim 0.04\left(n_i^{\prime}\omega _{pi}^{\prime ~2}/n_i\omega _{pe}^2\right)\left(
\Omega _ev_e/v_i\right)I/\left(1-I\right)$ rad s$^{-1}$. For $M=2$ the growth rate is
2 orders of magnitude lower. Thus for $n_i^{\prime}\sim 0.2 n_i$ the growth rates are
of order $10^{-2}-10^{-4}\times \left(v_e/v_i\right)I/\left(1-I\right)$ rad s$^{-1}$
for plasma parameters similar to those in Figure 8 ($\omega _{pe}=1.8\times 10^5$,
$\omega _{pi}=3\times 10^3$, $\Omega _e=1.8\times 10^4$ corresponding to $B=1$ mG).

The accelerated electron distribution function is given by the generalization
of equation (A16) in \citet{laming01}
\begin{equation}
f_e^{\prime}\left(v_{||}\right)={\left(-n_i^{\prime}kv_s\exp\left(-M^2/\alpha ^2\right)
/\alpha -n_i\omega\right)\over\sqrt{2\pi }kv_i^3}
{\omega _{pi}^2\over\omega _{pe}^2}\left[v_m-v_{||}+{\left(v_m^3-v_{||}^3\right)x^2
\over 3v_{\rm Ai}^2}\right],
\end{equation} where $v_m$ is a constant of integration, the maximum electron
velocity where $f_e^{\prime}\left(v_m\right) =0$.
Integrating over $v_{||}$ in the range $0\rightarrow v_m$, and making the
approximations $\omega << \sqrt{2}kv_i$, $\vec{k}\cdot\vec{U}=kv_s$ the
accelerated electron density is
\begin{equation}
n_e^{\prime}={n_i^{\prime}\over\sqrt{\pi }v_i^2}\left[-{M\over\alpha }+
{\alpha\left(M^2/\alpha ^2+1\right)^2\over
8M}\right]\exp\left(-M^2/\alpha^2\right)
{\omega _{pi}^2\over\omega _{pe}^2}\left({v_m^2\over 2}
+{v_m^4x^2\over 4v_{\rm Ai}^2}\right).
\end{equation}

For reference, the  accelerated electron density in
the opposite limit, $\omega >>
\sqrt{2}kv_i$ is given by \citep[][equation (A17)]{laming01}
\begin{equation}
n_e^{\prime}={n_i^{\prime}\over\sqrt{2\pi\rm e}v_i^2}
{\omega _{pi}^2\over\omega _{pe}^2}\left({v_m^2\over 2}
+{v_m^4x^2\over 4v_{\rm Ai}^2}\right),
\end{equation}
with $x\simeq\omega _{pi}/\omega _{pe}$.
Taking $\omega $ to be the lower hybrid frequency, equation (11) is valid for
$T_i>> 2.3\left(B/{\rm 1 mG}\right)^2\times 10^9/n_e$ K and equation (12) for
$T_i<< 2.3\left(B/{\rm 1 mG}\right)^2\times 10^9/n_e$ K. In between these limits Landau
damping by thermal ions is sufficiently strong to quench the instability.

To summarize, we have calculated wave dispersion relations and kinetic growth rates
in each of the limits $\omega >> \sqrt{2}kv_i$ and $\omega << \sqrt{2}kv_i$. Since
these growth rates are in general lower than the ion cyclotron frequency, $\Omega _i$,
we also calculate the group velocity of the waves, and by setting this equal to the
shock velocity find the angle to the magnetic field at which propagating waves can
remain in contact with the shock reflected ions and hence grow to large amplitudes
over essentially arbitrary periods of time. For these waves we calculate the growth
rates and hence densities of accelerated electrons.

We have revised the velocity range over accelerated electrons may exist from
$-v_m\rightarrow v_m$ in \citet{laming01}
to $0\rightarrow v_m$ due to the following consideration.
\citet{karney78} has derived a value for the maximum electric field above which
the wave becomes stochastic and ion heating also commences, given by
$E=B\left(\Omega _i/\omega\right)^{1/3}\omega /4/k_{\perp}=
B\left(\Omega _i/\omega\right)^{1/3}v_s/2$. The kinetic energy acquired by an
electron in time $t\sim 1/\Omega _i$, (the time available for electron acceleration
in the precursor before the shock hits) is $\left(m_i^2/2m_e\right)\left(E/B\right)^2
x^2/Z^2$, which for relevant values of $x$ and $\omega$ evaluates to energies
less than 1 keV for Cas A. This is valid for a true perpendicular shock. However it
is easy to see that an electron traveling along a magnetic field line inclined at
a small angle to the shock front (a quasi-perpendicular shock)
can easily stay ahead of the shock and be
accelerated to large energies, but may do so traveling in one direction
only.

The calculation of the accelerated electron distribution function assumes that
the electrons are accelerated purely by Landau damping of the lower hybrid wave,
which is likely to be the case for the non-relativistic electrons envisaged
here. However since lower hybrid waves exist with $\omega /k_{||}\sim c$,
electrons can in principle be accelerated to relativistic energies, if they can
be trapped in the wave generation region for long enough. \citet{mcclements97}
argue that this is indeed the case for quite a wide range of shock Alfv\'en
Mach numbers, based on the intensity at which lower hybrid waves saturate in
simulations and the resulting value of the momentum diffusion coefficient
for electrons, and that lower-hybrid waves may provide an electron injection
mechanism for subsequent Fermi shock acceleration to cosmic ray energies.

\section{Reverse Shock Ionization Structure}
\subsection{Initial Conditions and Basic Equations}
We model the hydrodynamics for the Cas A shell assuming
uniform density ejecta and ambient medium. We use expressions given by \citet{mckee95}
for the reverse shock radius and velocity with respect to the expanding ejecta,
$R_r$ and $\tilde{v_r}$, which connect between the early ejecta dominated, and
late Sedov-Taylor limiting behavior. The generalization of such models to
non-uniform densities is given in \citet{truelove99}. We choose parameters for the
hydrodynamic models of Cas A as follows. The blast wave velocity is inferred to be
5200 km s$^{-1}$ \citep{vink98,koralesky98}.
The expansion factor ($d\log R_b/d\log t$ where
$R_b$ is the radius of the blast wave) found by these authors to be $\eta = 0.64$
suggests that the remnant is slightly more than midway through the transition
between the ejecta dominated and Sedov-Taylor
dominated phases, where values of $\eta$ of 1.0 and 0.4 respectively would be
expected. Hence we expect $R_{ST}=2.23\left(M_e/M_{\sun}\right)^{1/3}n_0^{-1/3}$ pc and
$t_{ST}=209E_{51}^{-1/2}\left(M_e/M_{\sun}\right)^{5/6}n_0^{-1/3}$ years, the transition
blast wave radius and time \citep{mckee95}, to be similar to or perhaps slightly
smaller than the current values, 2.64 pc and 320 years respectively. In these
relations $n_0$ is the preshock hydrogen density,
$E_{51}$ is the explosion energy in units of $10^{51}$ ergs, and $M_e/M_{\sun}$ is
the ejecta mass in units of the solar mass.
The ionization age for this shock front is
determined to be in the range $10^{11} - 2\times 10^{11}$ cm$^{-3}$s,
\citep{vink96,favata97},which combined
with the age of 320 years gives a preshock electron density of 2.5 - 5.0 cm$^{-3}$,
assuming a shock compression of a factor of 4. Taking $R_{ST}=R_b=2.64$ pc gives
$M_e/M_{\sun}$ in the range 4-8. Assuming $t_{ST}=320$ years gives $E_{51}$ in the
range 2.4-4.8. From ASCA observations, \citet{vink96} find $M_e/M_{\sun}\sim 4$.
\citet{favata97} analyzing BeppoSAX spectra find $2<M_e/M_{\sun}<4$, with the uncertainty
stemming from different assumptions about the role of the non-thermal continuum. The
small ejecta mass is consistent with extensive mass loss as stellar winds by the
progenitor. Both references find O to be the dominant element composing the ejecta,
with rather little of the expected products of O burning evident, suggesting further
that these are confined in a compact stellar remnant. The emission measure quoted by
\citet{laming01} corresponds to 4$M_{\sun}$ of O (if $n_e=10$ cm$^{-3}$),
so in the following we will take
$E_{51}=2$, $M_e/M_{\sun}=4$, and $n_0=3$ cm$^{-3}$. This values are similar to
those considered by \citet{mochizuki99}, with the exception of a smaller $n_0$. It
appears that they chose $n_0$ consistent with the modeling of \citet{borkowski96},
who modeled the deceleration of the blast wave upon encountering a shell of
circumstellar material postulated to be at the interface of red and blue supergiant
presupernova winds, in order to match the measured radio expansion of $\sim 2000$
km s$^{-1}$. The more rapid expansion seen in X-rays \citep{vink98,koralesky98},
generally assumed to be more directly associated with the actual motion of the plasma,
suggests a lower circumstellar density as used here. \citet{borkowski96} do
demonstrate in their simulations one important feature relevant in this work, namely
the fact that the blast wave (and presumably also the reverse shock) split into
transmitted and reflected shocks upon encountering density contrasts, and that these
secondary shocks will propagate back across the presumed region of high magnetic
field near the contact discontinuity.

The blast wave of Cas A is observed at a radius of 160'', while the bright ring of
X-ray emission is at a radius of 110'' \citep{vink98,gotthelf01}. This suggests that the bright
ring is in fact emission from the ejecta heated by the reverse shock, since according
to the self-similar models of \citet{chevalier82}, for a blast wave radius of
160'', the radius of the contact discontinuity should be in the range 120''-140''.
Behind the reverse shock, the density $n_q$ of ions with charge $q$ is
given by
\begin{equation} {dn_q\over dt} =
n_e\left(C_{ion,q-1}n_{q-1}-C_{ion,q}n_q\right) +
n_e\left[\left(C_{rr,q+1} +C_{dr,q+1}\right)n_{q+1} -
\left(C_{rr,q}+ C_{dr,q}\right)n_q\right]
\label{eqn1}\end{equation}
where $C_{ion,q}, C_{rr,q}, C_{dr,q}$ are
the rates for electron impact ionization, radiative recombination and dielectronic
recombination respectively, out of the charge
state $q$. These rates are the same as those used in the recent ionization balance
calculations of \citet{mazzotta98}, using subroutines kindly supplied by
Dr P. Mazzotta (private communication 2000). The electron density $n_e$ is determined
from the condition that the plasma be electrically neutral. The ion and electron
temperatures, $T_i$ and $T_e$ are given by generalizations of equation (2) for the
case when many charge states of a particular element are present (see Appendix B)
\begin{equation} {dT_i\over dt}= -0.13n_e{\left(T_i-T_e\right)\over AT_e^{3/2}}
{\sum _qq^3n_q/\left(q+1\right)\over\sum _q n_q}
\end{equation} and
\begin{equation} {dT_e\over dt}= 0.13n_e{\left(T_i-T_e\right)\over AT_e^{3/2}}
{\sum _qq^2n_q/\left(q+1\right)\over\sum _q n_q}
-{T_e\over n_e}\left(dn_e\over dt\right)_{ion} - {2\over 3n_ek_{\rm B}}{dQ\over dt}.
\end{equation}
Here $A$ is the atomic mass of the ions in the plasma. The last term $dQ/dT$
represents plasma energy losses due to ionization and radiation. Radiation losses
are taken from \citet{summers78}. These authors tabulate the power emitted by
{\em each ion} of the elements C, O, Si, Ar, Fe and Mo as a function of temperature,
allowing an exact treatment of the ionization non-equilibrium aspect of the
power losses for these elements. Losses
for other elements can be readily obtained by interpolation along isoelectronic
sequences. At each time step $T_i$ and $T_e$ are modified
by a further factor $\exp\left(-4v_{ex}/3r\right)$ and $n_e$ and the $n_q$ by
$\exp\left(-2v_{ex}/r\right)$, coming from the adiabatic expansion of a spherical
shell of plasma with volume $V=4\pi r^2dr$ where $dr$ is held constant as $r$ increases.
In this approximation, $v_{ex}$, the
expansion velocity, which is constant for all ejecta, is
given by the lesser of $v_{ex}=R_r/t_r-0.75\tilde{v_r}$ and
$v_{ex}=0.75v_b$, with $t_r$ being the reverse shock
onset time, and $v_b$ being the blast wave velocity. This is a reasonable approximation
for the transition period between ejecta dominated and Sedov-Taylor evolution in a
supernova remnant \citep[see the ``hollow'' blast wave approximation in][]{ostriker88}.
More detailed work could use the self similar solutions appropriate
to the various regions and epochs of the remnant evolution
\citep[e.g.][]{chevalier82,hamilton84},
but even then the inevitable Rayleigh-Taylor instabilities are still difficult to treat.
The radius $r$ of the plasma element is given by $r=R_r +\int _{t_0}^tv_{ex}dt$, and we
neglect the fact that the rear side of Cas A is around 4pc further away from us than
the front side, and thus is observed about 13 years further back in time. Equations
(14) and (15) are integrated forwards in time from initial conditions given by the shock
jump conditions, equation (1), for the electron and ion temperatures and by the
hydrodynamic model discussed above for the densities. All ions are assumed to be singly
charged at the time of shock passage, and all ions of the same element are taken to
have the same temperature. Our results are insensitive to the exact initial
charge state distribution since we are mainly interested in what happens many years after
the initial postshock ionization has ceased.
In comparisons with the calculations of \citet{mochizuki99} for pure Fe ejecta we found
agreement to $\pm 10$\% or better for time dependence of the electron temperature
over the lifetime of Cas A.

\subsection{Ionization Structure with no Collisionless Electron Heating}
Figure 3 shows the evolution of the ion temperature (dashed lines) and the
electron temperature (solid lines) for O ejecta heated by reverse shock
encounters at various times in the evolution of Cas A. The highest electron temperature,
$T_e\simeq 2.7\times 10^7$ K
is found in plasma shocked about 150 years after the supernova.
Table 1 gives a
summary of the present day values of plasma parameters for different shells of the
ejecta shocked at different times after the initial explosion. In
Table 2, and illustrated in Figure 4 are similar temperature plots but for circumstellar
plasma with elemental composition dominated by N, He and H
(in mass ratios 0.02:0.49:0.49 respectively) shocked by the blast wave.
Here the highest electron temperature is $7.4\times 10^7$ K,
again for plasma shocked 150 years after explosion.
This value for the electron temperature
is consistent with that actually observed directly behind the blast wave (U. Hwang 2001,
private communication), but higher than that observed in spatially integrated spectra.
However we believe that the reverse
shocked ejecta is the dominant contributor to the observed X-ray emission. Besides the
morphological arguments given in the previous section, the blast wave as modeled
does not provide enough emission measure to account for the observed emission by
a factor of at least 4, remembering that this plasma being composed of less highly
charged ions radiates less bremsstrahlung per unit emission measure.
Although one could increase the circumstellar density and ejecta
mass in the model to match the observations, the explosion energy would also have to be
increased as well to to maintain the consistency with the observed blast wave velocity
and radius. A problem might still remain however with the likely overproduction of
O VIII emission from the dense ejecta cooling at temperatures around $10^6$ K, compared
with observations \citep[e.g.][]{bleeker01}. Only ejecta
which encountered the reverse shock early in the evolution of Cas A (i.e. at times
less than 100 years after explosion) has sufficient emission measure to account for
the observations.

High values of the ion temperatures in our simulation result simply from the choice
of the shock jump conditions, equation (1), as our initial conditions. Some observational
support for very high postshock ion temperature exists. \citet{ghavamian99} sees
broad H $\alpha$ profiles in optical spectra of Tycho, RCW 86 and the Cygnus Loop,
commensurate with the shock velocities derived in these remnants. Further,
\citet{raymond95} observing the NW limb of SN 1006 in the UV see broad line profiles
for C IV, N V, and O VI, indicating that the heavy ions have similar {\em velocity}
distributions to each other and to the protons, and temperatures proportional to their
masses consistent with equation (1).

So we proceed to place the dominant X-ray emission and the electron acceleration
in the ejecta. One immediate, but easily solved problem is that according to the
ionization balance calculations the temperature of the ejecta shocked at early times
should be below $10^6$ K, not $\sim 4\times 10^7$ K as observed. However,
if the hard X-rays are due to bremsstrahlung from suprathermal electrons, one
should expect just such a temperature deficit in order to be able to accommodate
the inevitable Coulomb heating of the ambient plasma due to the accelerated electrons.
Cas A emits $\sim 10^{35}$ ergs s$^{-1}$ above 10 keV. The ratio of radiated
bremsstrahlung power to that lost to Coulomb collisions is given approximately
by \citep[see e.g.][]{katz87}
$Zg\beta ^2\alpha _{\rm fs}/\left(3\sqrt{3}\log\Lambda\right)$ where $g\sim 1$ is the
bremsstrahlung Gaunt factor, $\beta =v/c$, and $\alpha _{\rm fs}$ is the fine
structure constant. For Cas A, $Z=8$, $\beta\sim 0.2$, and $\log\Lambda\simeq 40$
giving a ratio of $\sim 10^{-5}$. The nonthermal emission above 10 keV thus
requires an energy input of $\sim 10^{40}$ ergs s$^{-1}$, or a total energy
input integrated over the lifetime of the remnant of $\sim 10^{50}$ ergs.
The electron internal energy in an emission measure of $n_en_OV=3\times 10^{57}$
cm$^{-3}$ for $n_O\sim 2-5$ cm$^{-3}$ (corresponding to ejecta shocked 50 and 100
years after explosion) is $2.5n_eVk_{\rm B}T\sim 10^{49}$ ergs
for a temperature of $4\times 10^7$ K. However the energy required to maintain this
quantity of plasma at such temperature against energy losses from ionization,
radiation and adiabatic expansion since the initial reverse shock encounter
is more than this, $\sim 10^{50}$ ergs,
similar in order of magnitude to that available for heating from the accelerated
electrons.

\subsection{The Inclusion of Accelerated Electrons}
To investigate further the effect of the Coulomb heating of the ambient electrons
by the accelerated population, we plot in Figure 5 the time dependence of the
electron and ion temperatures behind the reverse shock under various heating conditions.
We assume a constant heat input per unit volume fixed at the present day value
in ejecta reverse shocked about 50 years after the explosion,
$\sim 4\times 10^{-16}$ ergs s$^{-1}$cm$^{-3}$, (see Appendix B.2 with $n_e=50$
and $n_e^{\prime}=2$). This heating is taken
to ``switch on''  at different times after reverse shock passage. These are
taken to be  150, 200, 250, and 300 years. A physical mechanism for such a
``switch on'' is suggested by the morphology of SN 1987A. The blast wave must encounter
a strong density gradient before reflected shocks can travel back towards the
contact discontinuity. Thus a certain amount of time may elapse while the blast
waves reaches this density gradient and the reflected shock propagates back. In SN 1987A
the relevant density gradient would be provided by the inner edge of the presumed
equatorial disk, the recombination radiation from which currently produces the inner
ring. In Cas A no real evidence for such an equatorial disk exists, but shock
reflection from a distribution of quasi-stationary flocculi is the main possibility.
>From Figure 5 it is apparent that electron heating beginning 200-250 years after
reverse shock passage (or after 250-300 years in the evolution of the supernova remnant,
i.e. between curves d and e)
will reproduce the current observed electron temperature of $\sim 4\times 10^7$ K.
Figure 1 illustrates this with spectra from
the BeppoSAX MECS \citep{boella97} and PDS \citep{frontera97} instruments acquired
on 1997 November 26, plotted against
a pure thermal bremsstrahlung spectrum at $4\times 10^7$ K
and a similar spectrum including an extra 4\% of electrons in an accelerated distribution
function. Also shown are seven data points from a rocket flight on
1968 December 5 \citep{gorenstein70}. These are shown in more detail in Figure 6
plotted against
pure thermal bremsstrahlung spectra corresponding to temperatures of $2\times 10^7$,
$3\times 10^7$ and $4\times 10^7$ K. From Figure 5 it is clear that on the basis
of the hydrodynamic models used in these calculations, the ambient electron temperature
should have increased between these two observations by $\sim 2\times 10^7$ K.
Depending on how one treats line emission and interstellar absorption (both omitted
from the theoretical spectra in Figure 5), such a temperature increase is marginally
consistent with the observations. The ionization age $n_et$ in such a case would be
$\sim 10^{11}$ cm$^{-3}$s$^{-1}$, a factor of $\sim 2$ lower than observations
\citep{vink96,favata97}. Curves b and c in Figure 5, representing electron acceleration
starting 150 or 200 years after explosion (100 or 150 years after reverse shock passage)
give values of $n_et$ more consistent with observations, but temperatures in the
range 5-6 keV. The heating mechanism envisaged here only occurs at perpendicular shocks,
and so not all the ejecta will be heated. Consequently a thermal conduction energy loss
from the heated regions might be expected, which would reduce the peak temperature
and reduce the degree of temperature increase that might be expected between the
observations of \citet{gorenstein70} and the present day. One interesting feature of
curves b and c is that the initial period of heating from $10^4$ K to $\sim 2\times 10^5$
K is very fast, but then slows abrubptly.
The ``plateau'' at this temperature is due to the strong peak in the
O radiative loss curve at this temperature, and the plasma temperature only starts
to rise again once it has expanded sufficiently to reduce the density so that the
collisionless heating can overcome the strong radiative losses.

We will now investigate in more detail exactly where in the Cas A ejecta a two stream
instability that provides lower hybrid wave electron heating may occur. In the picture
that we develop, the X-ray emitting shell of Cas A is filled with shocks propagating
in all directions as a result of forward or reverse shock interactions with
density inhomogeneities. In the hot ion limit, where $\omega <<\sqrt{2}kv_i$
the wave growth rates fall dramatically with increasing Mach number, so we can
reasonably assume that only shocks with the minimum Mach number to be supercritical
are important. For plasma $\beta$ (ratio of gas pressure to magnetic pressure) of 1
this Mach number is 1.7 for perpendicular shocks, falling to 1.35 for $\beta=2$
\citep{edmiston84}. A further simplifying assumption will be to assume that the
accelerated electrons heat
the ambient plasma to the observed temperature, $4\times 10^7$ K, and then that the
instability continues in marginal stability. This is a lesser criterion on the
electron energization process than the ambient electron heating discussed above.
In such conditions, the values of the parameters $I$ and $\left(k_{\perp}/2\right)
\partial I/\partial k_{\perp}$ in equation (A7) are $\sim 0.5$ and $\sim -0.25$
respectively. The accelerated
electron density is maximized for a given $k$
at $-M/\alpha \simeq 0.7$ from equation (11). This
is a lower value than is realizeable in practice, the minimum value being
$-M/\alpha\simeq 1$ for high $\beta$ plasmas and short wavelength waves. At marginal
stability, the electron plasma $\beta _e$ is given by $\beta _e= 8\pi n_ekT_e/B^2
=2\omega _{pe}^2v_e^2/\Omega _e^2/c^2 =2$. In heavy element plasma with highly charged
ions the electrons dominte the total plasma $\beta$,
so $M\simeq 1$. For $k=\omega _{pe}/c =\Omega _e/v_e$,
$-\alpha = 0.5 - 0.6$ (from Figure 7), so $-M/\alpha\simeq 2$.
Decreasing $\beta$, allowing larger $k$ and hence larger (i.e. more negative)
$\alpha$ also requires larger $M$, so $-M/\alpha \simeq2$ holds over the range of
reasonable values of $\beta$, $k$, and $\alpha$.
Assuming $n^{\prime}_i/n_i=0.2$ for the shock reflected ions, which is comparable
to that observed in situ at solar wind shocks \citep[e.g.][]{gedalin96}, the fraction
of electrons accelerated is given by $n_e^{\prime}/n_e=0.05/v_i^2$ where
$v_m=1.5\times 10^{10}$ cm s$^{-1}$ has been used in equation (11).
This corresponds to 70 atomic units found as the best match to BeppoSAX data
in \citet{laming01}.

The density of accelerated electrons may be increased over that
given by equation (11) if the energetic electrons are pitch angle scattered away
from the magnetic field direction.
This scattering mainly results from collisions with ions, while energy
loss is due to collisions with ambient electrons. The isotropic density of
accelerated electrons, $n_{e,iso}^{\prime}$ is given by
$n_{e,iso}^{\prime}\sim n_e^{\prime}\tau _{e,e}/\tau _{e,i}\sim 4n_e^{\prime}$. The
electron-electron and electron-ion scattering cross sections have the same
dependence on the energy of the accelerated electron, and so the isotropic
accelerated electrons will have the same energy distribution as those directed
along the magnetic field. This pitch angle scattering boosts the accelerated electron
density by a factor $\sim 5$ for the same reflected ion density. Hence an accelerated
electron fraction of 4\% may be possible in regions of the Cas A ejecta where
$v_i < 2.5\times 10^7$, i.e. $T_i < 6\times 10^7$ K.

>From Table 2 we see that ejecta
reverse shocked up to $\sim 100$ years after the initial explosion may support sufficient
electron acceleration to account for the observations.
In this context it is important to realize, however, that
ejecta shocked within 100 years of the explosion can have a low enough ion
temperature to justify using equation (12) for the accelerated electron density, which
would increase the accelerated electron fraction by around an order of magnitude
over that given by equation (11). In
general cool plasma instabilities are stronger than those in hot plasma, and so our
conclusion that electrons can be accelerated in ejecta shocked early in the
evolution of Cas A will be strengthened.

Hence our model for the electron acceleration to form the hard X-ray tail by
non-thermal bremsstrahlung and to provide heating of the ambient ejecta plasma to
$\sim 4\times 10^7$ K by Coulomb collisions requires the existence of shocked
ejecta that has cooled to give ion temperatures
well below $\sim 10^8$ K. This requires that secondary shocks passing through this
ejecta have velocities below $\sim 500 $ km s$^{-1}$ so as to avoid reheating the ions
up to temperatures where the instability weakens. This is actually quite likely, since
a shock reflected for example from the blast wave moving through the shocked
circumstellar medium will slow down considerably upon reaching the shell ejecta shell.

\section{Discussion}
We have previously argued \citep{laming98,laming00,laming01} that two stream instabilities
formed ahead of fast shock fronts in supernova remnants might more realistically be
modeled as kinetic rather than reactive instabilities. This is observationally
motivated, in that the degree of electron heating predicted by models based on
reactive instabilities \citep[e.g][]{cargill88} is not observed \citep{laming96,ghavamian99}.
A theoretical justification comes from the fact that high Mach number shocks are
turbulent and not steady state structures, and reflected ions consequently will not
form a ``beam'', but will inhabit a much wider portion of phase space, giving rise to
kinetic instabilities.
With their lower growth rates, these will generally allow insufficient wave growth
ahead of the shock before being overrun by the shock itself. In this paper we have
avoided this problem by focusing on lower hybrid waves formed ahead of the shock that
have group velocity away from the shock equal to the shock velocity itself, and hence
stay in contact with the shock reflected ions for sufficiently long times to allow
significant wave growth to occur.

In this paper we have found that the most efficient electron acceleration occurs for
the lowest Mach number shocks, just above the first critical Mach number, which is
where shock is said to become ``supercritical''. This denotes the onset of turbulence
in the shock structure, and ion reflection which becomes progressively more unsteady
and ``bursty'' in nature as the Mach number increases \citep[e.g.][]{tokar86}.
Hence although we considered
the lower hybrid instability excited by the secondary shocks in the Cas A shell to be
kinetic in nature since this appears to be a more robust assumption, the results of
our calculations suggest that in this case reactive instabilities could be quite
plausible also.
It is also worth remarking that the designation ``reactive'' or
``kinetic'' comes from the mathematical treatment of two different limits of the
Vlasov equation (see Appendix A),
and that physically, a continuum of instabilities exists between the
kinetic and reactive limits. The further investigation of these would require
numerical simulations. All instabilities in this paper have a minimum
wavector $k_{min}\sim\omega _{pe}/c$ arising from electromagnetic effects
and a maximum $k_{max}\sim\Omega _e/v_e$ from the requirement that the electron gyroradius
be smaller than the wavelength. In marginal stability $k_{min}\sim k_{max}$ allowing
us to estimate the magnetic field stength, $B=m_ev_e\omega _{pe}/e=5.9\omega _{pe}
\times 10^{-9}$, which evaluates to 1.9 mG for $\omega _{pe}=4\times 10^5$ rad s$^{-1}$
in the $t=50$ year ejecta.
>From Table 1 the ejecta shocked at 50 years after explosion occupies a volume
$\sim 2\times 10^{55}$ cm$^{-3}$, in order to give the stated emission measure.
The magnetic energy contained in this volume is $3\times 10^{48}$
ergs,  not inconsistent
with that deduced from the minimum energy synchrotron luminosity argument
\citep{longair94}.

These magnetic fields are assumed to be amplified from seed fields by the action of
a Rayleigh-Taylor instability \citep[though see][for another possibility]{lou94}.
Plausible evidence of this has recently been demonstrated
in imaging spectroscopy of Cas A by Chandra \citep{hughes00,hwang00}. In the SE
quadrant of Cas A in particular, emission in Fe L and K shell lines is observed
to be {\it outside} the emission from the lighter element Si, suggesting that the
Si burning zone in the progenitor (which produced the Fe) has overtaken the O burning
zone which produced the Si. This region in the SE is also where recent XMM-Newton
observations find the hardest continuum \citep{bleeker01}, though these authors comment
that the hard X-ray continuum emission actually is distributed quite uniformly over
the remnant, which argues in favor of an emission mechanism independent of the
magnetic field, i.e. bremsstrahlung, and not e.g. synchrotron emission. We have
previously argued that the secondary shocks that excite the lower hybrid waves
arise as the blast wave encounters density inhomogeneities in the circumstellar
medium producing reflected shocks as in \citet{borkowski96}. However another
possibility exists which is  perhaps more intimately associated with Fe emission and
Rayleigh-Taylor instability, viz. the reverse shock encountering Fe-Co-Ni bubbles
in the ejecta \citep{li93,borkowski00}. The simulation in the last reference
demonstrates that such a scenario would produce large amounts of turbulence and
vigorous mixing of Fe with the overlying ejecta. We also note in passing that
the idea of secondary shocks propagating through already shocked material, and also
being highly magnetic, has been successfully applied to ultraviolet spectra
of the bowshock in HH 47A by \citet{hartigan99}. In this case the strong magnetic
field arises as radiatively cooling gas is compressed by hotter material around it.
This is also possible but in our view unlikely for the regions of Cas A we discuss.

A common parameter quoted in models for electron heating is the fraction of the
shock energy that is deposited in the electrons. It is not possible to make
such a statement about the secondary shocks responsible for the electron
acceleration in our model of Cas A, but in terms of the blast wave energy an
estimate can be made. If the $10^{40}$ ergs s$^{-1}$ required to accelerate the
electrons comes from secondary shocks that originate from the blast wave, this energy
is to be compared with the energy $\sim 4\pi R_b^2v_sn_ik_{\rm B}T_i\simeq
10^{41}$ ergs s$^{-1}$ deposited by the blast wave in the shocked ions.
Neglecting energy going into cosmic ray ions, around 10\% of the blast wave energy
ends up in energized electrons to be compared with the $\sim 20$\%
estimated by \citep[e.g.][]{cargill88}.

The electron distribution function is given
by balancing the imaginary terms in the dielectric response, which give wave
growth due to shock reflected ions and Landau damping by the accelerated
electrons. We argue that secondary shocks maintain this electron distribution
against the inevitable losses due to Coloumb collisions, and that therefore
these shocks must pass through a plasma element on timescales significantly
shorter than those for Coloumb collisions, which are of order 2 - 40 years
for accelerated electrons of energy 10 - 70 keV in the ejecta shocked at early times.
The accelerated electron distribution will be
different from that modeled above if processes other than electron Landau
damping occur. \citet{mcclements97} argue that sufficient lower hybrid wave
intensity is likely to exist such that electrons may be accelerated by a stochastic
mechanism to mildly relativistic energies, but that this is more easily satisfied
at high Alfv\'en Mach number shocks.

\section{Conclusions}
In \citet{laming01} it was shown that in principle a distribution of electrons
accelerated by lower hybrid waves comprising about 4\% of the plasma electrons
produced a very good match to existing spectra of Cas A from BeppoSAX. This model
used cold plasma theory, assuming $\omega >>\sqrt{2}kv_i$. In this paper we have
extended the plasma theory to treat the opposite limit $\omega <<\sqrt{2}kv_i$,
and by performing calculations of the time dependent ionization balance within the
ejecta of Cas A, have identified regions where such wave generation may take place.

In ejecta shocked $\sim 150$ years or more since explosion, only the thermal plasma
instability ($\omega << \sqrt{2}kv_i$) may occur. However this instability appears to be
too weak to provide significant electron heating to give the hard X-ray emission.
In ejecta shocked up to $\sim 100$ years since
the explosion, either the cold or thermal plasma instabilities
($\omega >> \sqrt{2}kv_i$) can occur, depending on the ion temperature. The cold plasma
instability is more likely, and is sufficiently strong to explain the
putative hard X-ray bremsstrahlung.
Such electron heating and acceleration mechanisms may plausibly occur elsewhere, the
main requirement being a sufficently strong magnetic field.
We have concentrated our efforts on Cas A since it has the most conspicuous hard
X-ray emission, is well studied in other respects, and is thought to have undergone
relatively little elemental mixing \citep{johnston84} which renders the plasma
theory more tractable.

This work was supported by basic research funds of the Office of Naval Research
and has also made use of data obtained from the High Energy Astrophysics Science
Archive Research Center (HEASARC) provided by NASA's Goddard Space Flight Center.

\appendix
\section{Appendix: Lower Hybrid Waves}
\subsection{Reactive and Kinetic Instabilities}
In cold plasma, the dispersion relation for lower hybrid waves, omitting the
electromagnetic terms is
given by \citep[e.g.][]{vaisberg83,mcclements97,laming01}
\begin{equation}
 K^L=1+{\omega _{pe}^2\over\Omega _e^2}\sin ^2\theta
-{\omega _{pi}^2\over\omega ^2} -
 {\omega _{pe}^2\over\omega ^2}\cos ^2\theta =0
\end{equation}
where $\omega $ is the wave frequency, $\omega _{pe}$ and $\Omega _e$ are the
electron plasma frequency and gyrofrequency respectively, $\omega _{pi}$ is the
ion plasma frequency and $\theta $ is the angle between the direction of wave
propagation and the magnetic field vector. As $\sin\theta\rightarrow 1$ this gives
$\omega\rightarrow\Omega _{LH}$ where $\Omega _{LH}=\sqrt{\Omega _e\Omega_i}$,
the geometric mean of the electron and ion gyrofrequencies, known as the
lower hybrid frequency. In the presence of suprathermal ions
the dispersion relation is modified by the addition of the following term to the
expression for $K^L$;
\begin{equation}\label{eqn4}
 +{\omega ^{\prime 2}\over n^{\prime}_ik^2}\int
{\vec{k}\cdot\vec{v_i}\over\omega} {\vec{k} \over \omega
-\vec{k}\cdot\vec{v_i}}\cdot{\partial
f_i^{\prime}\over\partial\vec{v_i}} d^3\vec{v_i}
\end{equation}
where $\omega _{pi}^{\prime}$ $f^{\prime}\left(v_i\right)$ and $n^{\prime}$
are the plasma frequency, distribution function and
density of the suprathermal ions with velocity $v_i$, and
$\vec{k}$ is the wavevector with magnitude $k$.

If $\partial f^{\prime}/\partial v_i=0$
at $\omega=\vec{k}\cdot\vec{v_i}$ then the integral may be evaluated in a straightforward
manner. Taking the most extreme limit in this case we put $f^{\prime}=n^{\prime}\delta
\left(\vec{v_i}-\vec{v_b}\right)$ for a monoenergetic ion beam moving with velocity
$\vec{v_b}$, and the integral evaluates to $\omega _{pi}^2/\left(w-\vec{k}\cdot\vec{v_b}
\right)^2$. With the addition of this term to the expression for $K^L$ the
dispersion relation becomes a quartic equation analagous to that for the Buneman
instability, the complex roots of which (when present) give growing or damping waves.
Physically the wave growth is driven by the free energy in the ion beam as a whole.
Such instabilities are known as ``reactive'' instabilities and usually give very fast
growth rates.

In the opposite limit we make the replacement $\omega\rightarrow\omega +i\gamma$
assuming $\gamma << \omega$ and
evaluate the integral by going into the complex plane. This procedure is summarized by
the so-called Landau prescription
\begin{equation}
1/\left(\omega -\vec{k}\cdot\vec{v_i}\right)
\rightarrow P\left(1/\left(\omega -\vec{k}\cdot\vec{v_i}\right)\right)-i\pi\delta
\left(\omega-\vec{k}\cdot\vec{v_i}\right).
\end{equation}
Upon making this replacement, taking
imaginary parts and rearranging, an equation for the growth rate $\gamma$ in terms
of the integral
$\int\left(\vec{k}\cdot\partial f^{\prime}/\partial\vec{v_i}\right)\delta ^3\left(\omega
-\vec{k}\cdot\vec{v_i}\right)d^3\vec{v_i}$ results
\citep[see][for the full expression]{laming01}. In this limit the instability
is called ``kinetic'', and is physically due to stimulated Cerenkov emission of
plasma waves. For wave growth, $\partial f^{\prime}/\partial v_i >0$, or in other
words a population inversion must exist. This makes the analogy with quantum physics
clear, and indeed one can discuss the various wave-wave and wave-particle interactions
in terms of the appropriate Feynman diagrams.

Reactive and kinetic instabilities derive from the different mathematical treatments
of the term in the dispersion relation accounting for the suprathermal ions. Of course
in reality a continuum of instabilities exists between these two limits, but further
investigation of these requires numerical techniques.

\subsection{The Dispersion Relation in Thermal Plasma}
In finite temperature plasma, the general expression for the longitudinal part
of the dielectric tensor is \citep{melrose86}
\begin{eqnarray}
\nonumber K^L=&1+{4\pi q^2\over\omega ^2}\int\sum
_{s=-\infty}^{+\infty}{1\over\omega -s\Omega _e-k_{||}v_{||}}
{\left(s\Omega _e+k_{||}v_{||}\right)^2\over
k^2}J_s^2\left(k_{\perp}v_{\perp}\over \Omega _e\right)\left({\omega
-k_{||}v_{||}\over v_{\perp}}{\partial\over\partial p_{\perp}}
+k_{||}{\partial\over\partial
p_{||}}\right)f\left(\vec{p}\right)d^3\vec{p}\\ =&1-{\omega
_{pe}^2\over\omega ^2}{\omega m_e\over k^2k_{\rm B}T_e}\sum
_{s=-\infty}^{+\infty} \int _{-\infty}^{+\infty}{s^2\Omega _e ^2+k_{||}^2v_{||}^2+2s\Omega _e
k_{||}v_{||}\over\omega -s\Omega _e -k_{||}v_{||}} \left(m_e\over
2\pi k_{\rm B}T_e\right)^{3/2}\exp\left(-m_ev_{||}^2/2k_{\rm B}T_e\right)dv_{||}\\
\nonumber &\times\int _0^{+\infty}
J_s^2\left(k_{\perp}v_{\perp}\over\Omega _e\right)\exp\left(-m_ev_{\perp}^2/2k_{\rm B}T_e\right)
2\pi v_{\perp}dv_{\perp}.
\end{eqnarray}
Here $\omega $ is the wave frequency, $k_{||}$ and $k_{\perp}$ are the components
of the wavevector $k$ parallel and perpendicular to the magnetic field $B$ direction.
$\Omega _e $ is the electron cyclotron frequency, $v$ and $p$ are the electron velocity
and momentum, with subscripts having the same meaning as for the wavevector, and $J_s$
is the Bessel function of the first kind of order $s$.
Assuming $k_{||}v_{||}<<\omega -s\Omega _e$ in denominator, appropriate for lower-hybrid
waves, the expression simplifies to
\begin{eqnarray}
\nonumber K^L=&1-{\omega _{pe}^2\over\omega ^2}{1\over k^2}\sum _{s=-\infty}^{+\infty}
{\omega\over\omega -s\Omega _e}\left[s^2\Omega _e ^2\left(m_e\over k_{\rm B}T_e\right)^2
+k_{||}^2
\left(m_e\over k_{\rm B}T_e\right)\right]\\
&\times\int _0^{+\infty}
J_s^2\left(k_{\perp}v_{\perp}\over\Omega _e\right)\exp\left(-m_ev_{\perp}^2/2k_{\rm B}T_e\right)
2\pi v_{\perp}dv_{\perp}.
\end{eqnarray}
The term in $s=0$ is
\begin{equation}
K^L_{s=0}=-{\omega _{pe}^2\over\omega ^2}{k_{||}^2\over k^2}{m_e\over k_{\rm B}T_e}
\int _0^{+\infty}
J_0^2\left(k_{\perp}v_{\perp}\over\Omega _e\right)\exp\left(-m_ev_{\perp}^2/2k_{\rm B}T_e\right)
2\pi v_{\perp}dv_{\perp},
\end{equation}
while the terms in $s\ne 0$ are  evaluated by converting the $\sum _{-\infty}^{+\infty}
\rightarrow\sum _1^{+\infty}$. Using the standard property of the Bessel functions,
$1-J_0^2\left(z\right)=2\sum _{s=1}^{\infty}J_s^2\left(z\right)$,
and assuming $\omega << \Omega _e$ gives
\begin{equation}
\nonumber K^L_{\left|s\right|>0}={\omega _{pe}^2\over k ^2}
\left(m_e\over k_{\rm B}T_e\right)^2
\int _0^{+\infty}\left[1- J_0^2\left(k_{\perp}v_{\perp}\over\Omega _e\right)\right]
\exp\left(-m_ev_{\perp}^2/2k_{\rm B}T_e\right)
2\pi v_{\perp}dv_{\perp}.
\end{equation}
Including the term for thermal ions,
$\omega _{pi}^2/k^2v_i^2\left(1-\phi\left(\omega\over
\sqrt{2}kv_i\right)\right)$, where the usual plasma dispersion function is
$\phi\left(z\right)=-z/\sqrt{\pi}\int _{-\infty}
^{\infty}\exp\left(-t^2\right)/\left(t-z\right) dt$, a term unity for the vacuum,
and inserting an electromagnetic correction, $\omega _{pe}^2\rightarrow
\omega _{pe}^2/\left(1+\omega _{pe}^2/k^2c^2\right)$,
for the electron term involving $k_{||}$
\citep{begelman88,melrose86} the dispersion relation is
\begin{equation}
\omega ^2={\omega _{pe}^2\left(Ik_{||}^2/k^2\right)
/\left(1+\omega _{pe}^2/c^2k^2\right)\over
1+{\omega _{pi}^2\over k^2v_i^2}\left(1-\phi\right)+
{\omega _{pe}^2\over k^2v_e^2}\left(1-I\right)}
\end{equation}
where $I={m_e\over k_{\rm B}T_e}\int _0^{+\infty}  J_0^2\left(k_{\perp}v_{\perp}
\over\Omega _e\right)\exp\left(-m_ev_{\perp}^2
\over 2T\right)v_{\perp}dv_{\perp}$, $v_e^2=k_{\rm B}T_e/m_e$ and
$v_i^2=k_{\rm B}T_i/m_i$.

\subsection{Group Velocities}
Proceeding as in the cold plasma case, we differentiate equation (A8) with respect
to $k_{\perp}$ to derive an expression for the group velocity;
\begin{eqnarray}
\nonumber &{\partial\omega\over\partial k_{\perp}}=
{-\omega\over k}{k_{\perp}\over k}
+{\omega\omega _{pe\left(EM\right)}^2\over k^3c^2}{k_{\perp}\over k}
+{\omega\over 2I}{\partial I\over\partial k_{\perp}}\\
&+ {\omega\over 2}
\left[{2\omega _{pi}^2\over k^3v_i^2}{k_{\perp}\over k}\left(1-\phi
\right) +{\omega _{pi}^2\over k^2v_i^2}{\partial\phi\over\partial k_{\perp}}
+{2\omega _{pe}^2\over k^3v_e^2}{k_{\perp}\over k}\left(1-I\right)
+{\omega _{pe}^2\over k^2v_e^2}{\partial I\over\partial k_{\perp}}\right]
\left[1+{\omega _{pi}^2\over k^2v_i^2}\left(1-\phi\right)+{\omega _{pe}^2\over k^2v_e^2}
\left(1-I\right)\right]^{-1}
\end{eqnarray}
where $\omega _{pe\left(EM\right)}^2=\omega _{pe}^2/\left(1+\omega _{pe}^2/c^2k^2\right)$.

In the limit $\omega >> \sqrt{2}kv_i$, $\phi\rightarrow 1+k^2v_i^2/\omega ^2 +3k^4v_i^2
/\omega ^4+...$. We set $\partial\omega /\partial k_{\perp} =v_s =\alpha\omega
/k_{\perp}$ where $v_s$ is the shock velocity and $\alpha =1/\left(2\cos\beta\right)$
where $\beta$ is the angle between $\partial\omega /\vec{\partial k_{\perp}}$ and
$\vec{U}$, the bulk velocity of the shock reflected ion distribution.
With $x=k_{||}/k$ to lowest order in $\omega _{pi}^2/\omega _{pe}^2$,
\begin{eqnarray}
\nonumber &x^4\left[-{\omega _{pe\left(EM\right)}^4I\over k^2c^2}+
{\omega _{pe\left(EM\right)}^2I\over 1+\omega _{pe}^2\left(1-I\right)/k^2v_e^2}\right]+\\
\nonumber
&x^2\left[\omega _{pe\left(EM\right)}^2{k_{\perp}\over 2}{\partial I\over
\partial k_{\perp}} -\alpha\omega _{pe\left(EM\right)}^2I
+{\omega _{pe\left(EM\right)}^4I\over k^2c^2}+{\left(\omega _{pe\left(EM\right)}^4I/k^2v_e^2\right)
\left(k_{\perp}/2\right)\partial I/\partial k_{\perp}-
\omega _{pe\left(EM\right)}^2I
\over 1+\omega _{pe}^2\left(1-I\right)/k^2v_e^2}
\right]\\&
+\omega _{pi}^2 -\alpha\omega _{pi}^2+{\left(\omega _{pi}^2\omega _{pe}^2/k^2v_e^2\right)
\left(k_{\perp}/2\right)\partial I/\partial k_{\perp}\over
1+\omega _{pe}^2\left(1-I\right)/k^2v_e^2}=0.
\end{eqnarray}
This looks like a quadratic equation in $x^2$, but we must remember that $k^2=k_{\perp}^2
/\left(1-x^2\right)$, $\omega _{pe\left(EM\right)}\rightarrow\omega _{pe}$
and rewrite in terms of $k_{\perp}$, which gives a quartic equation
for $x^2$. However we are looking for solutions $x^2<<1$, for which both quartic and
quadratic equations give the same result to leading order in $\omega _{pi}^2/
\omega _{pe\left(EM\right)}^2$.
For $x^2<<1$, in the limit $k_{\perp}v_e/\Omega _e \rightarrow 0$, where
$I\simeq 1-k_{\perp}^2v_e^2/\Omega _e^2$ and $\left(k_{\perp}/2\right)
\partial I/\partial k_{\perp}\simeq I-1$ this has solution
\begin{equation}
x^2\simeq {\alpha\over -\alpha -1 + \omega _{pe\left(EM\right)}^2/k^2c^2}
\left(\omega _{pi}^2/ \omega _{pe\left(EM\right)}^2\right).
\end{equation}
In the cold plasma approximation, $\omega _{pe\left(EM\right)}^2/k^2c^2\rightarrow 0$
and $x^2=\omega _{pi}^2/ \omega _{pe}^2$ for $\alpha =-1/2$ (i.e. reflected ions
{\em returning} to the shock). This is the result given
previously \citep{mcclements97,laming01}. The electromagnetic correction however has
the effect of forcing the minimum wavevector to be $k=\omega _{pe}/c$.
We require $x^2>0$ which gives
$k^2> -\alpha\omega _{pe}^2/c^2/\left(\alpha +1\right)$. As $k_{\perp}v_e/\Omega _e$
increases from 0 the minimum wavevector decreases and reflected ions {\em leaving}
the shock can excite the necessary waves. For $k_{\perp}v_e/\Omega _e\sim 1$, so that
$I\simeq 0.5$ and $\left(k_{\perp}/2\right)
\partial I/\partial k_{\perp}\simeq -0.25$ with $\omega _{pe}>> kv_e$
\begin{equation}
x^2 = {1-2\alpha\over 1+\alpha - \omega _{pe\left(EM\right)}^2/k^2c^2} >0
\end{equation}
which requires $\omega _{pe\left(EM\right)}^2/k^2c^2-1 <\alpha <1/2$.
Figure 7 shows $x^2$ against
$k_{\perp}v_e/\Omega _e$ for $\alpha =1/2$ and $k^2=\omega _{pe}^2/c^2$ (bold curve),
taking $\omega _{pe}=4\times 10^5$, $\omega _{pi}=7\times 10^3$, and $\omega =560$
corresponding to $n_e=50$ and $n_i=6.25$ cm$^{-3}$ and a magnetic field of 1.9 mG.
The other curves show the loci of $x^2$ for $k^2$ multiplied by factors of 1/2, 2, and
4 (as labeled) from this value. Apart from the fact that it is now ions leaving
the shock that excite the waves, the value of $x^2$ is still essentially the same
as that found in the cold plasma case, i.e. $x^2\sim \omega _{pi}^2/\omega _{pe}^2$.

Now if $k\sim\omega _{pe}/c$, the approximation $\omega >>\sqrt{2}kv_i$ may be no
longer valid in regions of Cas A,
and we investigate the opposite limit of the plasma dispersion
function, $\omega << \sqrt{2}kv_i$. In between these two limits, where $\omega\sim
\sqrt{2}kv_i$ Landau damping by thermal ions will prevent any wave growth.
In the opposite limit $\phi\rightarrow\omega ^2/k^2v_i^2 +....$
and
\begin{equation}
x^2=1-{\alpha\left[1+\omega _{pi}^2/k^2v_i^2 +\left(1-I\right)\omega _{pe}^2/k^2v_e^2
\right]-\left(k_{\perp}/2I\right)\partial I/\partial k_{\perp}\left(1+\omega _{pi}^2
/k^2v_i^2 +\omega _{pe}^2/k^2v_e^2\right)\over \left[1+\omega _{pi}^2/k^2v_i^2 +
\left(1-I\right)\omega _{pe}^2/k^2v_e^2\right]\omega _{pe\left(EM\right)}^2/k^2c^2
-1},
\end{equation}
which can be rearranged to give
\begin{equation}
\alpha =\left(1-x^2\right)\left[{\omega _{pe}^2\over k^2c^2+\omega _{pe}^2}
-{1\over 1+\omega _{pi}^2/k^2v_i^2+\left(1-I\right)\omega _{pe}^2/k^2v_e^2}
+{k_{\perp}\over 2I}{\partial I\over\partial k_{\perp}}{1+\omega _{pi}^2/k^2v_i^2
+\omega _{pe}^2/k^2v_e^2 \over 1+\omega _{pi}^2/k^2v_i^2+\left(1-I\right)\omega _{pe}^2
/k^2v_e^2}\right].
\end{equation}
Again in the limit of low electron temperatures for $x^2<<1$
\begin{equation}
\alpha ={\omega _{pe}^2\over k^2c^2+\omega _{pe}^2}-{1+\omega _{pi}^2v_e^2/\Omega _e ^2v_i^2
+\omega _{pe}^2/\Omega ^2\over 1+\omega _{pe}^2/\Omega _e ^2+\omega _{pi}^2/k^2v_i^2}.
\end{equation}
For $\omega _{pe}^2 >>\Omega _e ^2$ this reduces to
$\alpha ={\omega _{pe}^2\over k^2c^2+\omega _{pe}^2}-1$ and satisfies $\alpha <-1/2$
for $k>\omega _{pe}/c$. In the opposite limit of high electron temperature,
$I\rightarrow 0$, $k_{\perp}\partial I/\partial k_{\perp}\rightarrow -I$, and
$\alpha =-1/\left(1+\omega _{pi}^2/k^2v_i^2\right)$ giving the more usual value
$k_{min}=\omega _{pi}/v_i$ \citep[see][]{laming01}. However in this limit the lower
hybrid waves do not exist, and so the appropriate wavevector for lower hybrid wave
growth is $k=\omega _{pe}/c$, arising from the electromagnetic correction. In Figure
7 we show the generalization of this result to arbitrary electron temperature.
The loci of $\alpha $ for $x^2=0$ are plotted against $k_{\perp}v_e/\Omega _e$,
taking the following
plasma parameters for Cas A, $\omega _{pe}=1.8\times 10^5$, $\omega _{pi}=3\times 10^3$,
and $\omega =300$, corresponding to $n_e=10$ and $n_i=1.25$ cm$^{-3}$, and a magnetic field
of 1 mG. The bold curve shows the value of $\alpha$ for $k=\omega _{pe}/c$, while the
curves show $\alpha\left(k_{\perp}v_e/\Omega _e\right)$ for values of $k$ successively
factors of 2 greater. The dotted line shows $\alpha =-1/2$, remembering that necessarily
$\alpha < -1/2$ for ions reflected back through the upstream plasma with relative velocity
approximately twice the shock velocity. In this limit wave propagating at a variety
of small cosines, $x$ to the magnetic field vector can stay in contact with shock reflected
ions. The wave direction that dominates will be that where the net growth rate is
maximized, as determined in section 2.

\section{Appendix: Collision Processes}
\subsection{Electron-Ion Coulomb Equilibration}
\citet{spitzer78} gives the timescale for an electron distribution to relax to a
Maxwellian as
\begin{equation}
t_{eq}\left(e,e\right)={3m_e^{1/2}\left(k_{\rm B}T_e\right)^{3/2}\over 4\pi ^{1/2}
n_ee^4\ln\Lambda }
\end{equation}
where $\Lambda$ is the so-called plasma parameter, the ratio of largest to smallest
impact parameters for collisions. In supernova remnants $\ln\Lambda\simeq 40$. The
equilibration time for ions
$t_{eq}\left(i,i\right)=t_{eq}\left(e,e\right)\sqrt{m_i/m_e}/Z_i^4$, and that for
electron-ion equilibration is $t_{eq}\left(e,i\right)=t_{eq}\left(e,e\right)m_i/m_e/Z_i^2$
where $Z_i$ is the ion charge. Accordingly we write
\begin{equation}
{d\Delta T\over dt}=-0.13Z^2n_e{\Delta T\over AT_e^{3/2}}
\end{equation}
which is equation (2), with $\Delta T=T_i-T_e$. We consider a fully ionized gas with
$n_e=Zn_i$ and
\begin{equation}
{d\over dt}\left(n_iT_i+n_eT_e\right)=n_i{dT_i\over dt}+ n_e{dT_e\over dt}=0.
\end{equation}
Solving these equations yields
\begin{eqnarray}
{dT_e\over dt}=0.13{Z^2n_e\over Z+1}{T_i-T_e\over AT_e^{3/2}}\\
{dT_i\over dt}=-0.13{Z^3n_e\over Z+1}{T_i-T_e\over AT_e^{3/2}}.
\end{eqnarray}
In deriving equations (14) and (15) these expressions are averaged over
the ion charge states in the plasma, and the expression for $dT_e/dt$ is
modified by the inclusion of terms accounting for the change in electron
density due to ionization, $-\left(T_e/n_e\right)\left(dn_e/dt\right)_{ion}$,
and radiative and ionization losses, $-\left(2/3n_ek_{\rm B}\right)dQ/dt$.
Recombinations, which reduce the electron density do not result in an increase
in the electron temperature in low density plasmas, since the energy of the
recombined electron is radiated away, rather than being shared with the other
plasma electrons as would be the case for three-body recombination in dense plasmas.

\subsection{Collisional Relaxation of Accelerated Electron Distribution}
The Boltzmann equation for the electron distribution function $f$ is
\citep[see e.g.][]{sturrock94}
\begin{equation}
{\partial f\over\partial t}=-\Gamma {\partial\over\partial v_r}\left(f{\partial H\over
\partial v_r}\right) +{\Gamma\over 2}{\partial ^2\over\partial v_r\partial v_s}
\left(f{\partial ^2G\over\partial v_r\partial v_s}\right)
\end{equation}
where $G=\int f\left|\vec{v}-\vec{v_1}\right|d^3\vec{v_1}$ and
$H=2\int f/\left|\vec{v}-\vec{v_1}\right|d^3\vec{v_1}$ are the Rosenbluth potentials
and $\Gamma= \left(4\pi e^4/m_e^2\right) \log\left(3\pi\Lambda\right)$
where $\Lambda$ is the usual
plasma parameter. For the collisional relaxation of a suprathermal electron distribution
$f$ with a background Maxwellian $f_1$ the Rosenbluth potentials take the forms
\begin{eqnarray}
{\partial ^2G\over\partial v_r\partial v_s}&=\int
f_1\left(v_1\right)\left[{\delta _{rs}
\over\left|\vec{v}-\vec{v_1}\right|}-{\left(v_s-v_{s1}\right)\left(v_r-v_{r1}\right)\over
\left|\vec{v}-\vec{v_1}\right|^3}\right]d^3\vec{v_1}\\ {\partial H\over\partial
v_r}&=2n_e\left[-{v_r\over v^3}\Phi\left(\beta v\right) + {1\over
v}{\partial\Phi\left(\beta v\right)\over\partial v}\right]
\end{eqnarray}
where $\Phi\left(\beta v\right)=2/\sqrt{\pi }\int _0^{\beta v}\exp\left(-x^2\right)dx$
with $\beta ^2=m_e/2k_{\rm B}T_e$. For a suprathermal electron distribution with
$v_{||}>>1/\beta $, $\Phi\left(\beta v\right)=1$. Working in one dimension then
$\partial ^2G/\partial v_{||}^2=0$ and $\partial H/\partial v_{||}=-2n_e/v_{||}^2$.
Hence
\begin{equation}
{\partial f\over\partial t}=2\Gamma n_e{\partial\over\partial v_{||}}\left(f\over v_{||}^2
\right)
\end{equation}
with $f\left(v_{||}\right)=n_e^{\prime}\left[v_m-v_{||}+\left(v_m^3-v_{||}^3\right)x^2/
3v_{{\rm A}i}^2\right]/\left(v_m^2/2+v_m^4x^2/4v_{{\rm A}i}^2\right)$, where $v_m$ is the
maximum electron velocity in the distribution, $v_{{\rm A}i}$
is the Alfv\'en speed and $x$ is the cosine of the angle between the magnetic
field vector and $\vec{k}$. In the cold plasma limit $\cos\theta=\omega _{pi}/\omega _{pe}$
so $v_{{\rm A}i}/x$ may be identified with the electron Alfv\'en speed. The energy loss
rate of the suprathermal electron distribution, which is equal in magnitude (but opposite
in sign) to the heating rate of
the ambient electrons, is given by
\begin{equation}
\int {1\over 2}m_ev_{||}^2{\partial f\over\partial t}dv_{||}=-{\Gamma n_en_e^{\prime}m_e\over
v_m^2/2 + v_m^4x^2/4v_{{\rm A}i}^2}\left[2\left(v_m+{v_m^3x^2\over 3v_{{\rm A}i}^2}\right)
\log {v_m\over v_{th}} -v_m +{v_m^3x^2\over 9v_{{\rm A}i}^2}\right].
\end{equation}
Taking the lower limit of integration to be $v_{th}=6\times 10^9$ cm s$^{-1}$, corresponding
to an electron energy of 10 keV, where the ambient and accelerated electron densities are
approximately equal,
$v_m\simeq 2.5v_{th}\simeq 1.5\times 10^{10}$ cm s$^{-1}$, and equation (B5) evaluates to an
energy input from accelerated electrons to the ambient plasma of
$4\times 10^{-18}n_en^{\prime}_e$ ergs cm$^{-3}$s$^{-1}$ for $v_{{\rm A}i}\sim xv_m$.

\figcaption{\label{fig1}Data from the BeppoSAX MECS and PDS
instruments with model bremsstrahlung spectra for a fully ionized
oxygen plasma. The softest model spectrum is pure thermal
bremsstrahlung. The harder spectrum has an additional 4\% of
electrons accelerated by lower hybrid waves to a maximum velocity of
$v_m=1.5\times 10^{10}$ cm s$^{-1}$ (see eq. 10). The seven data points in
bold print are taken from a rocket observation in 1968
\citep{gorenstein70} to illustrate the stability of at least the
thermal part on the spectrum.}
\figcaption{\label{fig2}Cartoon of the Cas A supernova remnant (not
to scale) showing the location of the blast wave or forward shock, the
reverse shock and the contact discontinuity between the reverse shocked
ejecta and the forward shocked circumstellar medium. Also shown schematically
are a few quasi-stationary flocculi (QSFs) ahead of the forward shock and fast moving
knots of ejecta (FMKs) ahead of the reverse shock. As the forward and reverse
shocks encounter these density contrasts, they split into transmitted and
reflected shocks. The reflected shocks can propagate back across the contact
discontinuity. Cas A has numerous knots and clumps of material, making it
likely that the entire shell is filled with these secondary shocks.}
\figcaption{\label{fig3}The
variation of ion temperature (dashed lines) and electron temperature (solid
lines) for reverse shocks in Cas A in pure oxygen ejecta, against time
since explosion. The evolution of the temperatures is plotted for ejecta
encountering the reverse shock at various times
(50, 60, 75, 100, 150, 200, and 250 years
after the initial explosion), and followed to 350 years after the explosion.
The present day corresponds to $t=321$ years, for an explosion date of 1680. The
most dense ejecta is that which encounters the reverse shock earliest, and in
this calculation undergoes radiative instability at around 120 years after explosion.}
\figcaption{\label{fig4}Similar to  Figure 1 but
for forward shocks in Cas A at 50, 150, and 250 years
since explosion. The elemental composition is now a mixture of N/He/H,
(in mass ratios 0.02:0.49:0.49) corresponding
to that observed in the quasi-stationary flocculi.
The electron temperature is again the solid line, the N ion temperatures are the dashed
lines, He ion temperatures the dotted lines, and the proton
temperatures the dash-dot lines. Very little ion-ion or electron-ion equilibration has
occurred.}
\figcaption{\label{fig5}Variation of ion (dashed
lines) and electron temperature (solid lines) in pure oxygen ejecta reverse
shocked at 50 years after the initial explosion, with electron
reheating by lower hybrid waves commencing a further 100, 150,
200, and 250 years since reverse shock passage, represented by the curves
labeled ``b'', ``c'', ``d'', and ``e'' respectively. Curve ``a'' shows the
initial heating and cooling following reverse shock passage, and is the same
as in Figure 3. The electron temperature is assumed to go no lower than
$10^4$ K, being maintained at this level by residual photoionization and conduction.
The electron heating is taken to be the same as the present day value, per unit
volume. The presently observed electron temperature is $\sim
4\times 10^7$ K, which is consistent with the curve with reheating
commencing between 250-300 years since explosion, or 200-250 years since
reverse shock passage.}
\figcaption{\label{fig6}Data
points from \citep{gorenstein70} plotted against pure thermal
bremsstrahlung spectra corresponding to temperatures of $2\times
10^7$, $3\times 10^7$, and $4\times 10^7$ K.}
\figcaption{\label{fig7}The variation of direction
cosine squared, $x^2=k_{||}^2/k^2$ against $k_{\perp}v_e/\Omega
_e$ for $\alpha=1/2$ taking $\omega _{pe}=4\times 10^5$, $\omega
_{pi}=7\times 10^3$, and $\omega =560$ corresponding to $n_e=50$
and $n_i=6.25$ cm$^{-3}$ and a magnetic field of 1.9 mG. The bold
curve gives the loci of $x^2$ for $k^2=\omega _{pe}^2/c^2$. The
other curves gives loci for $k^2$ equal to 1/2, 2, and 4 times
this value, as labeled. For $k_{\perp}v_e/\Omega _e\le 1$
direction cosines less than $\omega _{pi}/ \omega _{pe}$ are
found, in accordance with the cold plasma treament.}
\figcaption{\label{fig8}The variation of $\alpha$
with $k_{\perp}v_e/\Omega _e$ for various values of $k_{\perp}$.
The plasma parameters taken are $\omega _{pe}=1.8\times 10^5$,
$\omega _{pi}=3\times 10^3$, and $\omega =300$, corresponding to
$n_e=10$ and $n_i=1.25$ cm$^{-3}$, and a magnetic field of 1 mG.
The top curve gives $k_{\perp}=\omega _{pe}/c$. The successively
lower curves are for $k_{\perp}=a\omega _{pe}/c$ where $a=1.778$,
3.162, 10, and 100 respectively.}

\begin{deluxetable}{lrrrrrr}
\tablecaption{Reverse Shock Parameters from Ionization Structure
\label{table1}}
\tablewidth{0pt}
\tablehead{
\colhead{$t$ (years)}
& \colhead{$q$} &\colhead{$T_i$ (K)} &\colhead{$T_e$ (K)}
& \colhead{$n_e$ (cm$^{-3}$)}
& \colhead{$n_et$ (cm$^{-3}$s)}  &\colhead{approx. EM (cm$^{-3}$)}}

\startdata
50 & 0.726 & 1.00e4 & 1.00e4 & 51.8 &   & 5e57\\
100& 0.551 & 6.48e7 & 1.98e7 & 23.0 & 2.33e11  & 8e56\\
150& 0.432 & 5.36e8 & 2.73e7 & 11.8 & 6.89e10  & 3e56\\
200& 0.348 & 1.32e9 & 2.37e7 & 6.75 & 2.20e10  & 1e56\\
250& 0.286 & 2.13e9 & 1.71e7 & 3.83 & 6.51e9  & 6e55\\
300& 0.240 & 2.87e9 & 6.05e6 & 2.00 & 7.35e8  & 3e55\\
\enddata
\end{deluxetable}

\begin{deluxetable}{lrrrrrrrr}
\tablecaption{Forward Shock Parameters from Ionization Structure
\label{table2}}
\tablewidth{0pt}
\tablehead{
\colhead{$t$ (years)}
& \colhead{$T_N$ (K)}& \colhead{$T_{He}$ (K)}& \colhead{$T_H$ (K)}
&\colhead{$T_e$ (K)} & \colhead{$n_e$ (cm$^{-3}$)}
& \colhead{$n_et$ (cm$^{-3}$s)} &\colhead{approx. EM (cm$^{-3}$)}}

\startdata
50 & 6.30e9 & 1.83e9& 4.59e8 & 3.70e7 & 0.96 & 1.58e10  & 2e55\\
100& 1.02e10 & 2.98e9& 7.48e8 & 6.48e7 & 2.73 & 2.33e10  & 2e56\\
150& 1.13e10& 3.30e9& 8.27e8& 7.45e7 & 4.58 & 2.32e10 & 5e56\\
200& 1.05e10& 3.06e9& 7.67e8& 7.02e7 & 6.20 & 1.93e10  & 1e57\\
250& 8.99e9& 2.61e9& 6.53e8&  5.63e7 & 7.44 & 1.27e10  & 1.5e57\\
300& 7.46e9 & 2.14e9& 5.37e8& 3.12e7 & 8.56 & 4.17e9 & 2e57\\
\enddata

\end{deluxetable}

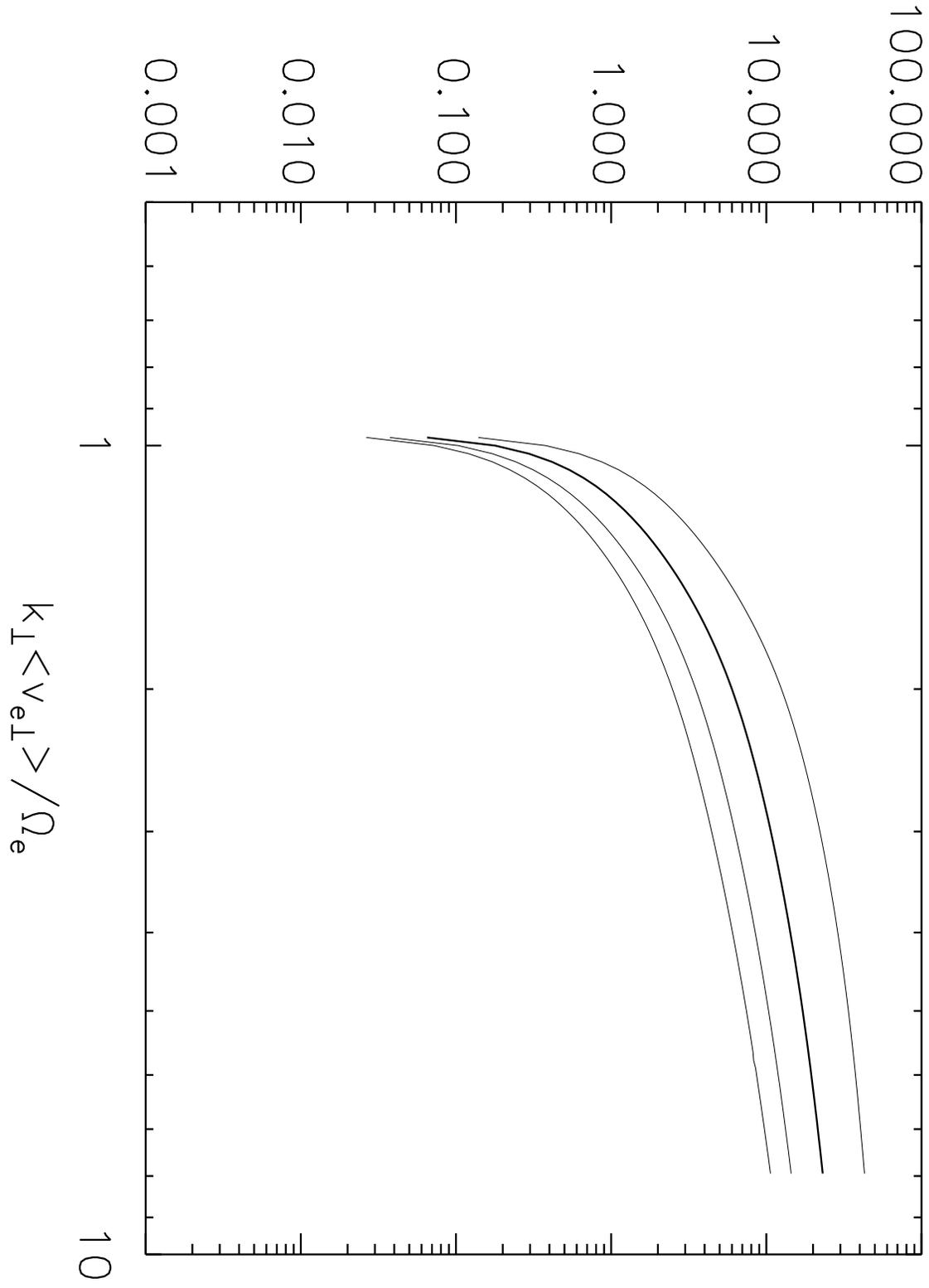
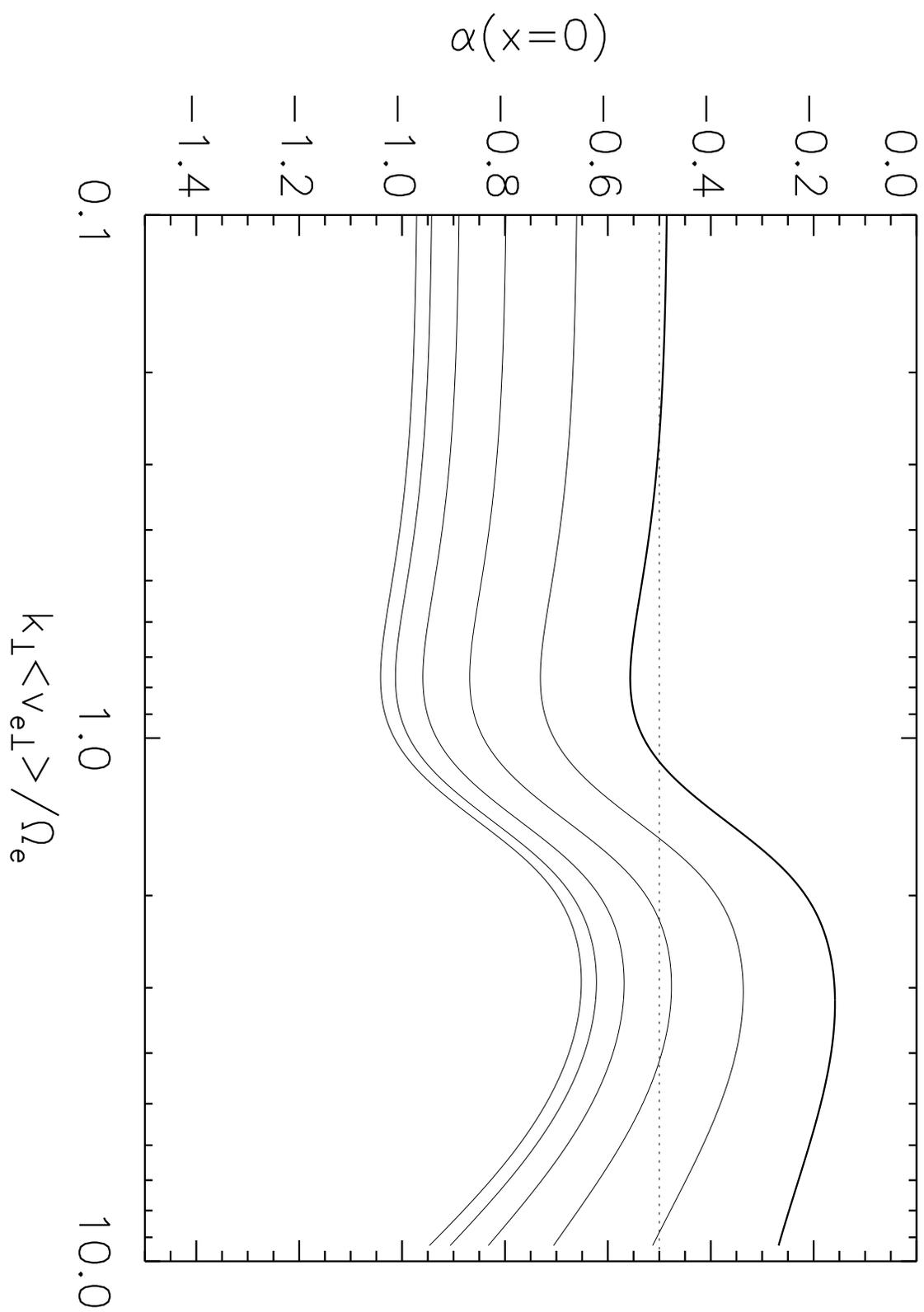
\end{document}